# THE URBAN OBSERVATORY: A MULTI-MODAL IMAGING PLATFORM FOR THE STUDY OF DYNAMICS IN COMPLEX URBAN SYSTEMS

Gregory Dobler[†,1234], Federica B. Bianco[1234], Mohit S. Sharma[4], Andreas Karpf[4], Julien Baur[4], Masoud Ghandehari[45], Jonathan Wurtele[6], Steven E. Koonin[45]

*August 30, 2019*

**Abstract:** We describe an "Urban Observatory" facility designed for the study of complex urban systems via persistent, synoptic, and granular imaging of dynamical processes in cities. An initial deployment of the facility has been demonstrated in New York City and consists of a suite of imaging systems – both broadband and hyperspectral – sensitive to wavelengths from the visible (~400 nm) to the infrared (~13 micron) operating at cadences of ~0.01 – 30 Hz (characteristically ~0.1 Hz). Much like an astronomical survey, the facility generates a large imaging catalog from which we have extracted observables (e.g., time-dependent brightnesses, spectra, temperatures, chemical species, etc.), collecting them in a parallel source catalog. We have demonstrated that, in addition to the urban science of cities as systems, these data are applicable to a myriad of domain-specific scientific inquiries related to urban functioning including energy consumption and end use, environmental impacts of cities, and patterns of life and public health. We show that an Urban Observatory facility of this type has the potential to improve both a city's operations and the quality of life of its inhabitants.

## 1. Introduction

With millions of interacting people and hundreds of governing agencies, urban environments are the largest, most dynamic, and most complex human systems on earth. Some 80% of the US population (and 50% of the global population) now live in cities [1], and within those urban boundaries are a multitude of interactions between the three fundamental components of urban systems: the inhabitants, the natural environment, and the built environment. The study of cities as complex systems is not new [2-4]. Research in urban planning [5,6], engineering [7,8], transportation [9,10], etc., all have a rich history of quantifying city life at multiple scales and with networked interactions [11,12]. The application of that work addresses everything from quality of life [13-15] to public health [16,17] to sustainability and resilience [18-20]. However, there are two recent revolutions that are leading to a dramatic change in the way we understand complexity in urban environments: the systematic collection, digitization, and curation of vast quantities of urban records data [21-23] and the development of computational techniques that make it possible to jointly analyze large, disparate data sources.

---
[†] gdobler@udel.edu
[1] Biden School of Public Policy and Administration, University of Delaware
[2] Department of Physics and Astronomy, University of Delaware
[3] Data Science Institute, University of Delaware
[4] Center for Urban Science and Progress, New York University
[5] Civil and Urban Engineering, Tandon School of Engineering, New York University
[6] Department of Physics, University of California, Berkeley

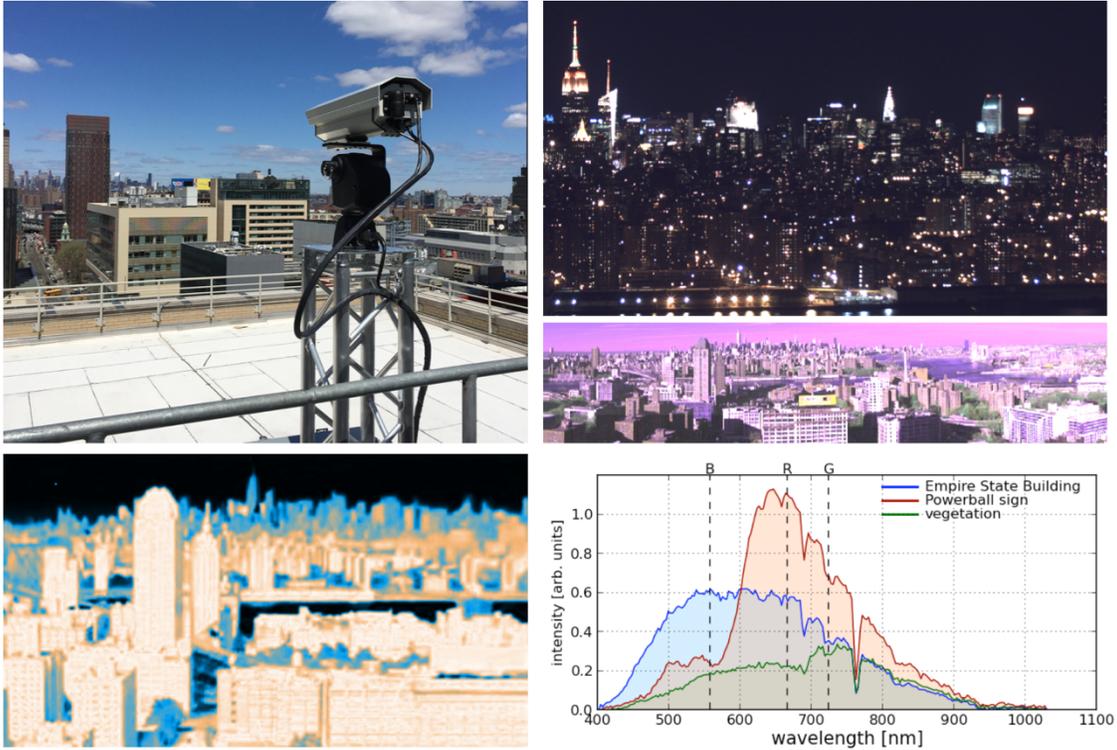

**Figure 1.** *Top left*: an example of one of our Urban Observatory (UO) deployments of a Visible Near-Infrared (VNIR) Hyperspectral camera. The camera itself is encased in a weatherized housing and sits atop a pan/tilt, both of which are connected to (and controlled by) our backed computing infrastructure. *Upper right*: An example broadband nighttime image of Midtown Manhattan and the Lower East Side captured by that vantage point [49]. *Lower left*: An example infrared image captured from the same vantage point. *Middle and lower right*: An example VNIR hyperspectral scan of the same scene shown in false color. The *middle* image is integrated over all wavelengths while the *lower* panel shows three spectra from the scene [50].

An important field of study for complex urban environments is that of remote sensing [24-28] and the associated data collection from overhead satellite platforms at a variety of wavelengths [29,30]. This satellite data is both large in volume and rich in information content, and its analysis has been used to correlate remote observables with land use characteristics [31], energy consumption [32-36], light pollution [37-39], poverty [40-42], and urban growth [43-45] (among many others). Characteristically, these overhead platforms provide spatial resolution of ~10s of meters and temporal resolution ~1 day or more.[1] However, at present, this resolution is insufficient to study dynamical interactions in cities on the time scale of minutes, and the overhead vantage point – while ideal for total spatial coverage – only captures photons that are emitted upward providing an incomplete picture for several science cases (e.g., the ecological and public health impacts of light pollution [37,46-48]).

In this paper, we describe an "Urban Observatory" (UO) for the study of dynamical processes in complex urban systems. As we outline in §2, our initial deployment of this UO in New York City (NYC) consists of side-facing imaging systems mounted atop tall buildings, providing a synoptic view of city skylines (see Figure 1) [49]. The spatial resolution of UO instrumentation is sufficiently granular to segment the imaging at the sub-building level, and (inspired by astronomical observatories operating in survey mode such as the *Sloan Digital Sky Survey* [51] or PanSTARRS [52]) the system operates persistently in order to observe multiple temporal scales: minutes, hourly, daily, weekly, monthly, annual, etc. In §3, we give an in-depth description of the range of urban phenomenology that is accessible via this system. This includes work done to date on three key components of complex urban systems and urban metabolism more generally: energy use in cities, environmental impacts, and human factors. In §4 we conclude with what we envision to be important future deployments of UO facilities in other cities.

---

[1] though recent advances in satellite technology are pushing towards increased granularity on both fronts.

## 2. Instrumentation

The core UO platform consists of a suite of imaging systems mounted atop tall buildings at distance of 1-4 kilometers from the area of study, along with associated computing hardware located both at the device (where minimal edge computations are performed) and at a central server location. The former consists of small mini-computers (Raspberry PIs or similar), while the latter is a multi-purpose backend infrastructure that

1. drives the imaging devices remotely;
2. pulls and queues data transfer from the deployed devices;
3. performs signal processing, computer vision, and machine learning tasks on the images and associated products; and
4. holds and serves imaging and source databases associated with the data collection and analysis.

This backend infrastructure/architecture is described in detail in Appendix A.

While we envision that the UO will ultimately carry out observations across the full (available) electro-magnetic spectrum at multiple time scales, the current deployments – which yield the science content described in §3 – consist of the following modalities.

### 2.1 Broadband Visible

The first deployed camera system was a three color (RGB), 8 Megapixel USB camera with a 35mm lens mounted atop a tall building in Brooklyn, NY with a panoramic view of the Manhattan skyline (Figure 1). The instrument was set to a fixed pointing, enclosed in a weatherized housing, and triggered to acquire images via a co-located laptop in a weatherized casing. The initial image capture rate was $f = 0.1$ Hz [49]. Current deployments have incorporated a pan/tilt mechanism for increased field-of-view (FOV) via multiple pointings, mini-PCs for triggering and data transfer, and ethernet controlled instruments of two types: 20 Megapixel cameras (sampling at $f = 0.1$ Hz) and DSLRs operating in video mode at a sampling frequency of $f \approx 30$ Hz.

### 2.2 Broadband Infrared

Our current broadband infrared (IR) devices are FLIR A310/320 cameras with a pixel resolution of 320x240, wavelength range of 7.5–13 micron, and temperature sensitivity of ±2° C. As with our visible wavelength imaging, our initial IR deployment was encased in a weatherized housing, had a fixed pointing, and operated at $f = 0.1$ Hz (Figure 1 shows an example image), while subsequent deployments incorporate a pan/tilt mechanism for increased FOV.

### 2.3 Visible and Near-IR Hyperspectral

In addition to our broadband imaging devices, we have deployed hyperspectral imagers operating at visible and near-infrared (VNIR)

wavelengths. These instruments are single-slit spectrographs: the aperture is a vertical slit while a diffraction grating behind the slit generates the wavelength information. They are mounted atop pan/tilt mechanisms that provide horizontal information as the detector is exposed at ≈30 frames per second during the pan. The wavelength range is 0.4–1.0 micron with an average bandwidth of 0.75 nm resulting in ~850 spectral channels [50]. Scans are captured at cadences of ~$10^{-3}$ Hz. An example image and associated spectra are shown in Figure 1.

*2.4 Longwave IR Hyperspectral*

In April of 2015, the UO carried out a test deployment of a Long Wave IR (LWIR) hyperspectral camera. This actively cooled instrument was sensitive to 7.5–13.5 micron in 128 spectral channels and was operated in survey mode at roughly f ~ 0.01 Hz, using the same panning mechanism described for the VNIR camera above. The deployment was done in collaboration with the *Aerospace Corporation* from whom the UO rented the equipment for an 8-day observational campaign [53].

*2.5 Privacy Protections*

Given the nature of our observational platform, it is of highest importance that the UO ensure appropriate privacy protections for inhabitants of cities [54]. This imperative is not unique to the UO nor is it restricted to imaging modalities; all observational platforms (whether acoustic sensors, air quality sensors, traffic sensors, etc.) *must* enact strict limits on the type and content of the data they collect – and the analysis procedures applied to that data – to ensure the privacy of individuals.

The UO's privacy protection policies consist of four core components

1. No personally identifiable information (PII) [55] is collected.

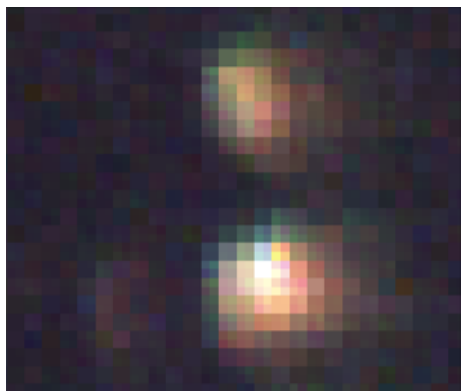

**Figure 2.** A zoomed-in example of two of the closest sources in the upper right panel of Figure 1 (figure reproduced from [49]). The resolution of all UO imaging is sufficiently low that no interior information is visible or detectable.

2. All imaging is strictly limited in pixel resolution so that building interiors cannot be seen (see Figure 2).
3. All analysis is aggregated and de-identified.
4. An individual source cannot be tracked across multiple modalities to infer PII.

In addition, it is UO policy that data and analyses are not voluntarily shared with enforcement agencies (e.g., law enforcement, environmental regulatory enforcement, etc.).

## 3. Urban Science and Domains

The instrumentation and associated operational modes described in §2 enable a wide variety of domain science inquiries related to the dynamics of urban systems from sub-second to yearly time-scales. Below, we describe several core aspects of city functioning that we are exploring through the application of computer vision, image processing, machine learning, and astronomical analysis techniques to the UO data streams. These avenues of study can be largely grouped into three categories: Energy, Environment, and Human Factors. These represent topical areas that are closely associated with the fundamental components of urban systems. The science focus of the UO is to develop a deeper understanding of cities through observational determination of the dynamical interplay between these areas.

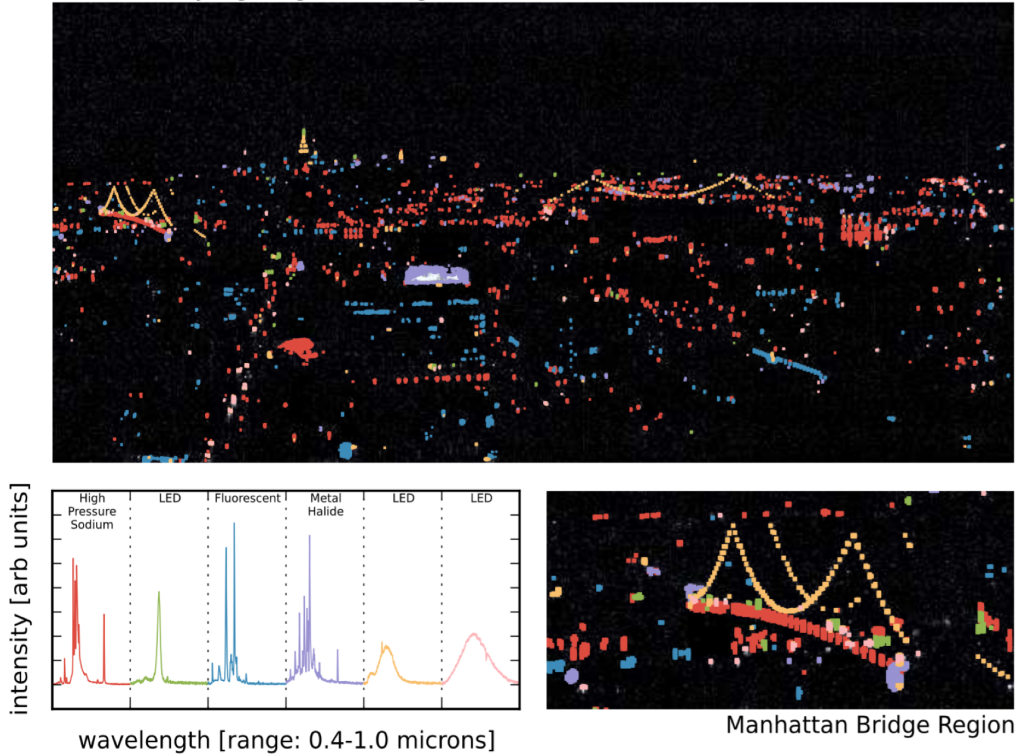

**Figure 3.** A night time hyperspectral scan of the scene in Figure 1 (reproduced from [50]). The *top* panel shows each source in the scene color coded by the lighting type determined from the spectra (*lower left*). The *lower right* panel shows a zoomed in view of the Manhattan Bridge region with LED necklace lights, High Pressure Sodium street lights, LED car headlights, and Metal Halide spotlights.

We note that, regardless of imaging modality (e.g., high frequency video or low frequency hyperspectral), the segmentation of each image into individual buildings by geolocating each pixel in the image, is an essential first step in all of the topics described below. Furthermore, this "building ID" step allows us to fuse imaging data from the UO with publicly available records data which can enable lines of inquiry that are otherwise impossible. The details of our methods for image segmentation and building ID are given in Appendix B.

*3.1 Energy*

Urban energy use – including electricity, natural gas, heating oil, etc. – drives city functioning and is at the heart of urban science questions related to resource consumption and distribution, infrastructure performance and maintenance, and future urban planning. Furthermore, it serves as the primary source of cities' impacts on the environment [56]. The UO's imaging capabilities with their multiple spatial and temporal scales allow for numerous unique ways of quantifying urban energy use characteristics from remote vantage points.

*3.1.1 Remote Energy Monitoring*

Lighting variability in the cities can be observed via broadband visible imaging at 0.1 Hz (10 seconds between images). In [49] we showed that aggregated lighting variability in NYC displays a diurnal pattern of use (on/off behavior) that roughly repeats night-to-night. This variability encodes occupancy characteristics that serve as strong correlates to total energy consumption [57]. We find that using these imaging data as an input to a convolutional neural network (CNN) trained on ground truth consumption data results in a 10-15% improvement on network-level electricity consumption prediction models trained on temperature and humidity data alone.

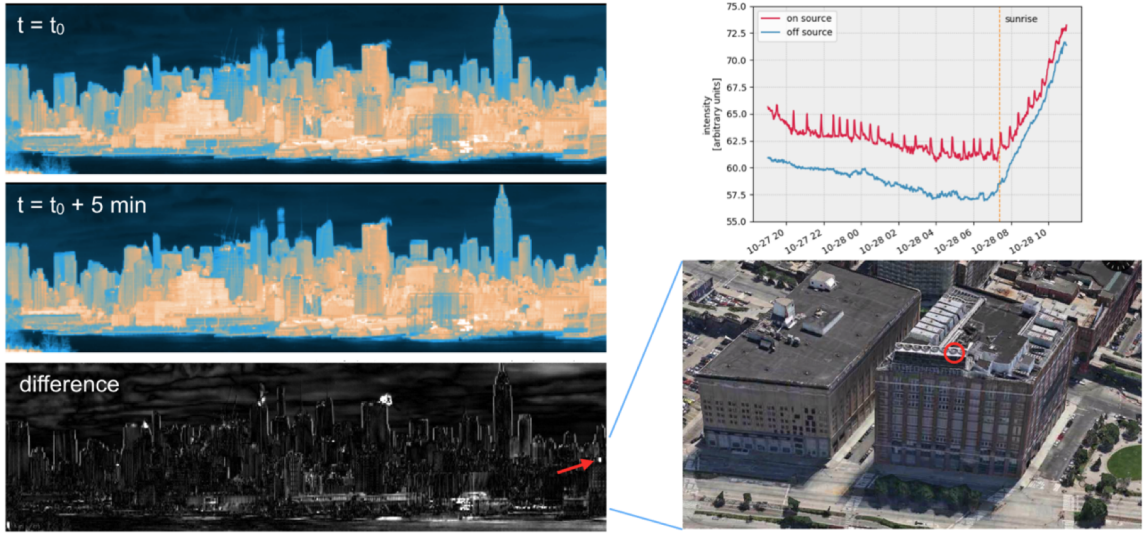

**Figure 4.** *Left:* two IR images separated by 5 minutes and the difference between the two. The difference image shows both steam plumes as well HVAC vents (*lower* right) that heat up and cool down with building energy use. *Top right:* the time-dependent signature of a similar source (in red) compared to the building facade (in blue) over one night shows the cadence of the building's heating/cooling system.

### 3.1.2 Lighting Technologies and End-Use

The VNIR hyperspectral instrumentation described in §2.3 has sufficient spectral resolution and photometric sensitivity to identify characteristic emission features in nighttime lighting [58,59] at a distance of several kilometers. In [50,60] we showed that this information can be used to determine the lighting type of individual sources via comparison to spectra measured in the lab and developed Template Activated Partition (TAP) clustering to identify lighting types that were not previously cataloged. An example of the technique is shown in Figure 3.

Lighting technology identification has several important energy uses including identification of targets for energy efficient lighting upgrades by practitioners, empirical determination of end-use characteristics by consumers [61], and measuring rates of adoption of modern lighting technologies by developing cities [62].

### 3.1.2 Grid Stability and Phase

The 60hz frequency of the AC mains is reflected in a sinusoidal oscillation of flux in the lights connected to the mains for a subset of lighting technologies, e.g., incandescent, traditionally-ballasted halogen, fluorescent, etc. In [63] we showed that observing city lights at high temporal frequencies allows us to monitor the grid frequency at city scale. However, since persistent observations at frequencies of 100s of Hz is not feasible (e.g., due to data transfer and storage limitations), in [63] we used a liquid crystal shutter operating at ~119.75 Hz to generate an ~0.25 Hz beat frequency that is observable with cadences of

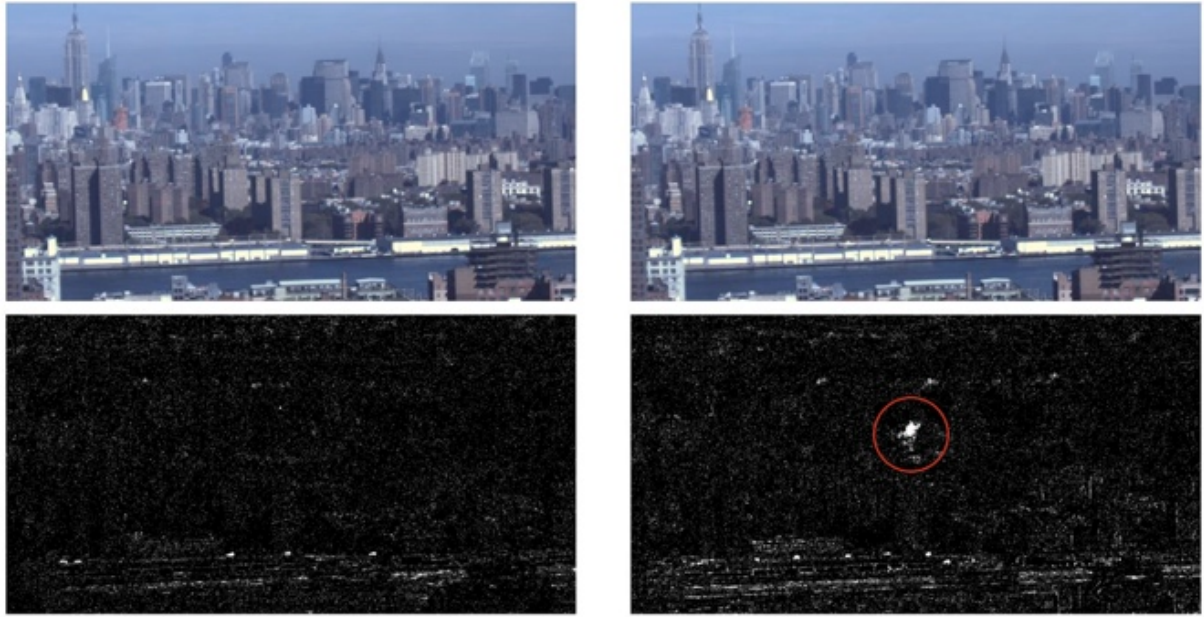

**Figure 5.** *Top panels*: Two broadband visible daytime images separated by one minute. *Bottom panels*: the same two images, but with the time-independent background removed. The application of foreground/background separation techniques clearly reveals a soot plume (circled in red) that has been ejected from one of the buildings in the scene.

~several Hz. As we described, this capability allows for monitoring of the stability of the grid across an urban distribution network, detection of relative phase of sources, and (with sufficiently accurate observations) potentially also transient phase shifts due to load at the unit level.

### 3.1.3 Building Thermography at Scale

In Figure 4 we demonstrate the use of time-dependent broadband infrared imaging to study thermographic envelopes of buildings in the city skyline. In particular, not only are efficiency characteristics such as heat leaks and thermal couplings detectable [64,65], but the figure also shows that individual HVAC vent duty cycles can also be seen at a distance as well.

As with the broadband visible wavelength imaging, this type of source variability can serve as an input to energy consumption models trained on consumption data. In addition, Building Management System operations and heating/cooling efficiency can be measured at scale across thousands of buildings from a single IR sensor.

### 3.2 Environment

Energy use in cities (and urban metabolism more broadly) generates byproducts of that use that have significant environmental impact. These impacts have local effects (e.g., degraded air quality leading to breathing-related health outcomes [66-70]), regional effects (e.g., fine-particle air pollution of regions surrounding cities [71]), and – due to the physical and population size of large cities – global effects (e.g., greenhouse emissions and reduced biodiversity [72,73]). The UO instrumentation extends traditional remote sensing of environment by satellites to increased spatial and temporal resolution (albeit with decreased geo-spatial coverage) to allow for the study of *dynamical* detection of environmental impacts of cities on sub-minute timescales.

### 3.2.1 Soot Plumes and Steam Venting

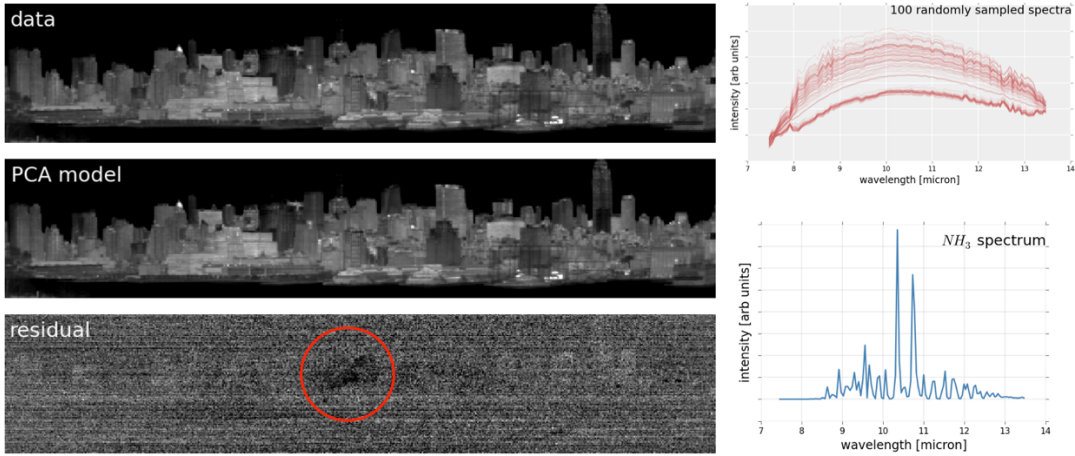

**Figure 6.** *Left*: A Principle Component Analysis (PCA) model fit to an Long Wave Infrared LWIR hyperspectral image. The PCA model (middle) was reconstructed from the first 10 principle spectral components of spectra in the scene (*top right*) and the model at 10.3 micron was subtracted from the raw data (top) to produce the 10.3 micron residual in the *bottom left* panel. The result is a clear, extended Ammonium plume, which has a 10.3 micron absorption feature as shown in the *lower right*.

As buildings in urban environments burn oil for heat, they produce soot plumes that are dispersed through the ambient air as they are advected away from the source by local winds. Such plumes are responsible for ~75% of greenhouse gas production in NYC [74], and are detectable in UO visible wavelength observations at cadences of 0.1 Hz. Figure 5 shows an example of one such detection. In the raw imaging, the dark, very low surface brightness of the plume makes it extremely difficult to detect directly from these raw data. However, foreground/background separation techniques [75,76] reveal the plume clearly towards the center of the image. In subsequent images, the plume is blown to the right as it disperses. In addition to soot plumes, the venting of steam from building heating and cooling systems is also visible.

The tracking of plumes has significant potential for not only monitoring the total number of plumes produced and the resulting effects on local air quality, but their motion can also be used as tracers of urban winds, informing studies of air flows through complex urban terrains (including simulations of such effects; e.g., [77]). The plume in Figure 5 is a particularly striking example, however most plumes are quite difficult to detect through classical computer vision techniques due to the complex urban background, time-dependent shadowing, low source brightness, and amorphous nature of the plumes. We have recently found that applications of regional convolutional neural networks (R-CNN) [78-80] can be tuned to detect such plumes and we have developed end-to-end tracking systems for application to these type of data [81].

*3.2.2 Remote Speciation of Pollution Plumes*

A variety of molecular compounds have strong absorption and emission features in the 7.5–13.5 micron range of the LWIR instrument described in §2.4. Over the course of that observational campaign we showed in [53] that numerous molecular compounds, including Ammonia, Methane, Acetone, Freon-22, $CO_2$, etc., could be identified in plumes emitted from buildings along the NYC skyline. In Figure 6, we show a simple detection of an Ammonia plume using a Principle Component Analysis (PCA) decomposition of a single data cube produced by the LWIR instrument. The various PCA components capture blackbody radiation, atmospheric (including water vapor) effects, and instrumental artifacts. The PCA model for each pixel, when subtracted from the raw data reveals a spatially localized deficit (i.e., a plume) in the 10.35 micron residual that is due to the absorption line of Ammonia at that wavelength. This technology has significant applications for both environmental studies of cities as well as emergency management and tracking of toxic materials release.

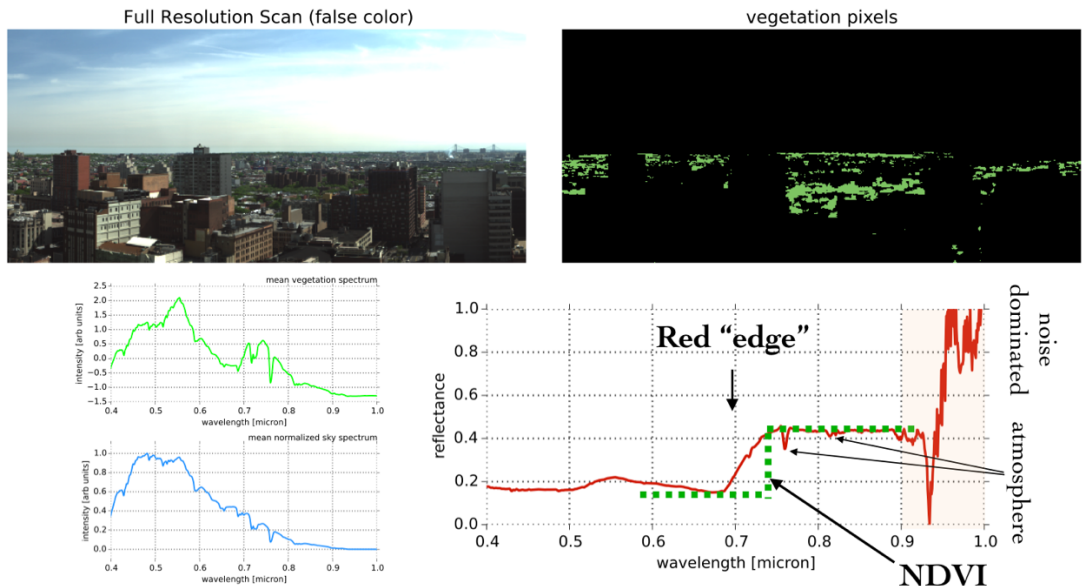

**Figure 7.** *Top left*: a false color representation of a VNIR scan of Brooklyn, NY during the daytime. *Top right*: the vegetation pixels in the scene identified by clustering-based image segmentation. *Bottom left*: the mean vegetation spectrum and mean sky spectrum in the scene. *Bottom right*: the ratio of those two giving plant reflectance as a function of wavelength. The Normalized Difference Vegetation Index (and other vegetative health indices) is easily determined given the spectral resolution and sensitivity of our VNIR instrument.

*3.2.3 Urban Vegetative Health*

The UO's daytime VNIR observations have sufficient sensitivity to directly measure chlorophyll features in the reflectance spectra of urban vegetation [82]. In particular, Figure 7 shows the characteristic "red edge" at ~700 nm that can be used to determine vegetative health indices such as the Normalized Difference Vegetation Index (NDVI) [83].

Furthermore, because of the extremely high spectral resolution and sensitivity, as well as the persistent nature of the UO's observational strategy, new metrics on vegetative health can be developed, low signal-to-noise effects such as solar-induced fluorescence [84] can be measured, and short timescale response to local air quality can be determined to high precision.

*3.2.3 Ecological Impacts of Light Pollution*

It is well known that city lighting has detrimental impacts on local ecologies including affecting migratory avian behavior [85]. These effects are both regional [86] and highly local (i.e., individual light installations [87]). In collaboration with the *New York City Audubo*n bird conservancy, we have deployed visible wavelength cameras acquiring images at 0.1 Hz to detect time-dependent lighting in lower Manhattan. By combining this data with regional NEXRAD radar scans [88] we are able to measure the ecological impacts from urban lighting on migratory bird densities at scales of ~100s of meters and at time scales of minutes.

*3.3 Human Factors*

Urban functioning is fundamentally driven by human decision making. Infrastructure use, transportation choice and their management, economic transactions, etc., all have, at their core, a human element that drives observable patterns of activity. This micro-behavior aggregates to macro-scale patterns of life with diurnal, weekly, monthly, and annual rhythms that can be detected by deployed sensing platforms like the UO.

*3.3.1 Patterns of Lighting Activity and Circadian Phase*

In [49] we showed that aggregate lighting activity in NYC displays clear diurnal patterns that repeat day after day and week after week. These patterns (on/off transitions of individual light sources) differ for residential versus commercial buildings and – as noted above – can serve as proxies for occupancy characteristics of buildings. In addition, these patterns for residential buildings correlate with the circadian behavior of the population as shown in Figure 8 [89]. Variations of these patterns with ambient lighting intensity can quantify the effects of light pollution on public health [48,90].

Interestingly, in [49] we also showed that, while the aggregate patterns were strictly repeating – albeit with different behavior on weekends versus weekdays – a given source does *not* strictly repeat from one day to the next (nor from one day to one week later). This type of UO data directly address the micro/macro behavioral properties of population and the scale of the transition between the two.

*3.3.2 Technology Adoption and Rebound*

Technological choice, and in particular the correlation of choice with socio-economic and demographic characteristics of the population, is an important indicator of population dynamics. In the energy sector, choice is an end-use characteristic studied most commonly by surveys of users [91]. By combining the lighting patterns described in §3.3.1 with the lighting technology identification described in §3.1.2, our UO platform is ideally suited to not only quantify choice as a function of population characteristics via fusion (as described in Appendix B) of imaging data with census data, but also allows for an empirical measurement of the amplitude of the rebound effect [91] in which the benefits of energy efficiency are offset by increased use stemming from a decreased economic incentive of curtailing use.

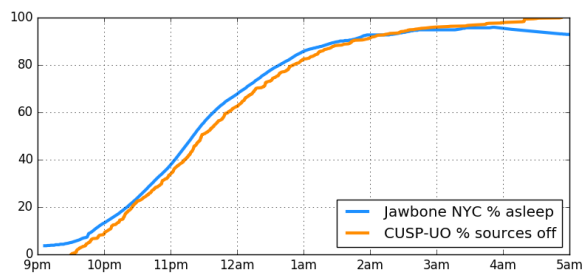

**Figure 8.** An overlay of the percentage of sources "off" after 9:30pm on a single Monday night (from UO broadband visible night time imaging of the scene in Figure 1) with the fraction of Jawbone$^{TM}$ users asleep in NYC on Monday March 31st 2014. There is a clear correspondence between our observations of lighting variability and the aggregate circadian patterns of NYC.

**4. Conclusions**

Modern cities are systems with tremendous complexity, and the functioning and behavioral properties of those systems has significant implications for their current and future environmental impacts as well as the quality of life of the inhabitants. It is through a detailed analysis of the three fundamental components of urban systems (the inhabitants, the natural environment, and the built environment) that one can uncover the dynamics that govern urban behavior. Measuring those dynamical interactions of urban systems requires high spatial and temporal resolution, with sufficient coverage to generate a representative sample of that system as a whole. We have presented an observational platform for the collection of data that can provide inputs to machine learning, computer vision, image processing, and astronomical analysis techniques that extract information relevant to the functioning of cities. Our realization of that platform in New York City is the creation of the Urban Observatory facility, consisting of imaging systems sensitive to visible and infrared wavelengths (both broadband and hyperspectral), with an operational mode that is persistent, synoptic, and granular. The urban science and domain topics that this data can address are broad ranging from energy use and its environmental impacts to patterns of life and public health. As the technology develops, deployment of similar Urban Observatories to cities of various sizes, local environments, and localities will enable a comprehensive and rich comparative study of diverse cities, deepening our core understanding of complex urban systems.

**Acknowledgements** GD's and MSS's work has been supported by a Complex Systems Scholar Award from the *James S. McDonnell Foundation* (JSMF). FBB, GD, JB, and MSS have been partially supported by Department of Energy (DOE) ARPA-E IDEAS grant. Imaging deployments have been supported by the JSMF, DOE ARPA-E, and Leon Levy Foundation. We thank the New York City Audubon for their partnership and collaboration.

## Appendix A: Backend Infrastructure

The deployed instruments and their associated operational modes described in §2 require a flexible and robust backend computing infrastructure to collect, store, and analyze the large volume of data generated by the UO. This infrastructure consists of both hardware and software components that, taken together, operate continuously to yield persistent observations. Our backend infrastructure consists of the following core components with associated functionality.

*Camera control devices* – Each imaging device is equipped with a mini-computer that opens a direct link with the camera itself. This machine is tasked with communicating directly with the camera and issues image acquisition commands. In certain instances, this computer can also be used to perform edge computations including compression or sub-sampling of the data. Acquired data may be saved temporarily on disk on this machine for buffered transfer back through the gateway server, or be written directly to bulk data storage.

*Gateway server* – The main communications hub between our computing and data storage servers and the deployed instrumentation is a gateway server that issues scheduled commands to the edge mini computers. This hub is also responsible for the pull (from the deployment) and push (to the bulk data storage) functionality for the data acquisition as well as the gateway for remote connections of UO users to interact with our databases.

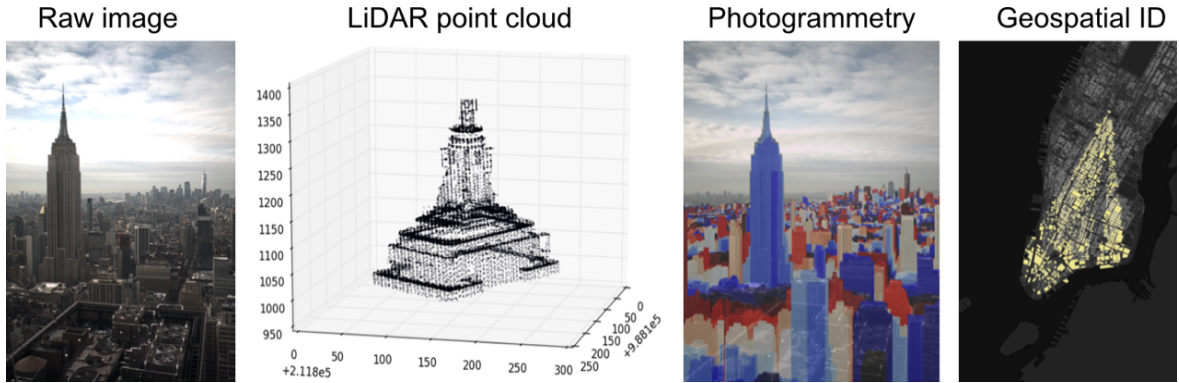

**Figure B1.** *Left*: a broadband visible image from a UO deployment in midtown Manhattan. *Center-left*: a LiDAR point cloud of the top of Empire State Building. Photogrammetric techniques allow us to use LiDAR to identify the buildings observed by each pixel (*center-right*) given the building tax-lot footprints from a publicly available data base (*right*). With these building IDs, we can integrate geospatial information with image data.

*Bulk data storage* – At full operational capacity, a UO site (consisting of a broadband visible camera operating at 0.1 Hz, broadband infrared camera operating at 0.1 Hz, a DSLR operating in video mode, and a VNIR hyperspectral camera operating at $10^{-3}$ Hz) acquires roughly 2–3 TB per day. This data rate necessitates not only careful data buffering and transfer protocols to minimize packet loss from the remote devices, but also a large bulk data storage with an appropriate catalog for the imaging data. This ~PB-scale storage server is directly connected to our computing servers for computational speed. This storage server also hosts parallel source catalogs that store information extracted from the data.

*Computing server* – Our main computing server that is used to process UO data consists of a dedicated >100 core machine that is primarily tasked with processing pipelines including: registration, image correction, source extraction, etc. We have designed our own custom containerized interface that seamlessly allows UO users to interact with the data while background data processing and cataloging tasks operate continuously.

*GPU mini-cluster* – Several of the data processing tasks described in §3 require the building and training of machine learning models with large numbers of parameters including convolutional neural networks. For these tasks, we use a GPU mini-cluster that is directly connected to our main computing server and which is continuously fed streaming input data from which objects and object features are extracted.

## Appendix B: Data Fusion

To maximize the utility of our UO facility, it is important that we be able to integrate the massive imaging data sets that we generate with available geo-spatial data including publicly available data such as census data, building-level energy consumption, fuel types, external sensor data (e.g., air quality sensors or radar data), etc.

Our data fusion utilizes publicly available LiDAR data from NYC [92] to locate the 3-space coordinate that is covered by each pixel. Specifically, using the collinearity equations, we project the topographic LiDAR point cloud into the 2D-image plane (given the position, pitch, yaw, roll, and focus of the camera) and, for each pixel, choose the closest LiDAR point. Since the LiDAR resolution is ~1 foot, there are pixel lines-of-sight for which the nearest LiDAR point is behind the nearest surface in that direction, and so we make a "no overhang" approximation and assign a given pixel the same 3-space coordinate as the pixel above it if the pixel above it is found to be closer to the camera. Finally, we use the publicly available MapPLUTO (Primary Land Use Tax-lot Output) data that contains the geospatial footprints for each building to associate the x,y components of the 3-space coordinate of a pixel with a given building footprint. Thus we are able to tag each pixel as "observing" a given building (see Figure B1). Additional geo-spatial data at coarser resolution (e.g., census tracts, utility distribution zones, etc.) can be associated as well.